\documentclass{amsart}
\usepackage{amsmath}
\usepackage{amssymb}
\DeclareMathOperator{\vecc}{vec}

\usepackage{graphicx}

\usepackage[sorting=anyt,citestyle=authoryear]{biblatex}
\allowdisplaybreaks[1]
\usepackage{thmtools,thm-restate}

\newcommand{\Det}{\operatorname{Det}}

\theoremstyle{definition}

\theoremstyle{remark}

\usepackage{algorithm}
\usepackage{algorithmicx}
\usepackage{algpseudocode}

\numberwithin{equation}{section}

\begin{document}

\title{Note on the geodesic Monte Carlo}

\author{Andrew Holbrook}
\address{Department of Statistics, UC Irvine}
\email{aholbroo@uci.edu}
\thanks{I would like to thank Simon Byrne for helpful discussion.  This work was sponsored by NSF grant DMS-1622490.}



\date{}



\begin{abstract}
Geodesic Monte Carlo (gMC) is a powerful algorithm for Bayesian inference on non-Euclidean manifolds. The original gMC algorithm was cleverly derived in terms of its progenitor, the Riemannian manifold Hamiltonian Monte Carlo (RMHMC). Here, it is shown that alternative and theoretically simpler derivations are available in which the original algorithm is a special case of two general classes of algorithms characterized by non-trivial mass matrices.  The proposed derivations work entirely in embedding coordinates and thus clarify the original algorithm as applied to manifolds embedded in Euclidean space.
\end{abstract}

\maketitle

\section{Introduction}

Bayesian inference is hard.  Bayesian inference on non-Euclidean manifolds is harder.  Prior to the publication of \textcite{byrne2013geodesic}, a statistician required great ingenuity to compute the posterior distribution for any model with non-Euclidean parameter space, and the algorithmic details might change significantly depending on the prior, the likelihood, and the constraints implied by the non-Euclidean geometry.  A good example of this approach is found in \textcite{hoff2009simulation}, where the posterior distribution over the Stiefel manifold of orthonormal matrices is computed by way of column-at-a-time Gibbs updates that rely on model specifications.

It is preferable, rather, that the same algorithm work for many different kinds of models.  This is one of the strengths of Hamiltonian Monte Carlo \parencite{duane1987hybrid} and its Riemannian extension, RMHMC \parencite{girolami2011riemann}, which augments the posterior distribution $\pi(q)$ by the random Gaussian momentum $p \sim N\big(0,G(q)\big)$, where $G(q)$ is the metric tensor pertaining to the Riemannian manifold over which the model is defined.  RMHMC simulates from the posterior distribution by simulating the augmented canonical distribution with Hamiltonian
\begin{align}\label{RMHMC}
    H(q,p) = U(q) + K(q,p) \propto - \log \pi(q) + \frac{1}{2} \log |G(q)| + \frac{1}{2} p^TG(q)^{-1}p \, ,
\end{align}
i.e., $U(q)$ is the negative log-posterior and $K(q,p)$ is the negative logarithm of the probability density function of Gaussian momentum $p$. Since the kinetic energy is not separable in $q$ and $p$, the system is not integrable using Euler's method, so, in most cases, implicit integration methods are required \parencite{girolami2011riemann}. However, \textcite{byrne2013geodesic} point out that, for certain manifolds with known geodesics, it is beneficial to split the Hamiltonian into two parts and simulate the two systems iteratively. Here, the first Hamiltonian $H^{[1]}=- \log \pi(q) + \frac{1}{2} \log |G(q)|$ renders the equations
\begin{align*}
    \dot{q} &= 0 \\
    \dot{p} &= \nabla_q \big(\log \pi(q) - \frac{1}{2} \log |G(q)| \big)\, ,
\end{align*}
and, crucially, the second Hamiltonian $H^{[2]}=\frac{1}{2} p^TG(q)^{-1}p$ renders the geodesic dynamics for the Riemannian metric's Levi-Civita connection. Thus, the entire system may be simulated by iterating between $(1)$ perturbing the momentum and $(2)$ advancing along the manifold geodesics.

\section{gMC on embedded manifolds}
\textcite{byrne2013geodesic} extends the RMHMC formalism to posterior inference on manifolds embedded in Euclidean space. In the following, this extension is referred to as the embedding geodesic Monte Carlo (egMC). To maintain the RMHMC formalism, the authors begin by considering the inference problem on the \emph{intrinsic} manifold, where the Hausdorff measure
\begin{align*}
    \mathcal{H}^d(dq) =\sqrt{|G(q)|} \, \lambda^d(dq) \, ,
\end{align*}
and not the Lebesgue measure $\lambda^d(dq)$, is the base measure with respect to which the posterior distribution is defined\footnote{Whereas the ensuing derivation is extremely clever, it is unfortunate that it relies on an intrinsic conception of the inference problem, which, we will argue, causes confusion when the object of interest is \emph{a priori} defined using the Euclidean embedding coordinates.}. Here, the RMHMC Hamiltonian \eqref{RMHMC} may be written
\begin{align*}
    H(q,p) = - \log \pi_\mathcal{H}(q) + \frac{1}{2} p^TG(q)^{-1}p \, ,
\end{align*}
for
\begin{align*}
    \log \pi_\mathcal{H}(q) =  \log \pi(q) - \frac{1}{2} \log |G(q)|
\end{align*}
the log-posterior with respect to the Hausdorff base measure. Now, a clever change of variables occurs using an \emph{isometric embedding} as a tool. An isometric embedding of a manifold $\mathcal{Q}$ into Euclidean space is a map $x: \mathcal{Q} \rightarrow \mathbb{R}^d$ satisfying
\begin{align*}
    G_{ij}(q) = \sum_{l=1}^d \frac{\partial x_l}{\partial q^i}(q) \frac{\partial x_l}{\partial q^j}(q) \, ,\quad \mbox{or}\quad G(q)=J_x(q)^TJ_x(q) 
\end{align*}
for $J_x(q)$ the Jacobian of the map $x$ evaluated at $q\in \mathcal{Q}$. \textcite{byrne2013geodesic} use the isometric embedding to make gMC practical on certain manifolds. This is accomplished by the change of variables $(q,p)\mapsto \big(x(q),Mp\big)$, with
\begin{align*}
   M=J_x(q)\big(J_x(q)^TJ_x(q)\big)^{-1}= J_x(q)G(q)^{-1} \, .
\end{align*}
If $v=Mp$, then the Hamiltonian $H(q,p)$ becomes (\cite{byrne2013geodesic}, Equation (9))
\begin{align}\label{gMCham}
    H(x,v) &= - \log \pi_\mathcal{H}(x) + \frac{1}{2} v^T \Pi_q v  \\ \nonumber
    &=  - \log \pi_\mathcal{H}(x) + \frac{1}{2} v^T v \,  
\end{align}
 for $\Pi_q$ the projection matrix of the tangent space of the embedded manifold (at point $q$) conceived of as a subspace of the ambient Euclidean space.  The authors point out that ``the target density $\pi_\mathcal{H}(x)$ is still defined with respect to the Hausdorff measure of the manifold, and so no additional log-Jacobian term is introduced,'' and invite the reader to 
 
 \begin{quote}
 [n]ote that by working entirely in the embedded space, we completely avoid the coordinate system $q$ and the related problems where no single global coordinate system exists. The Riemannian metric $G$ only appears in the Jacobian determinant term of the density: in certain examples, this can also be removed, for example by specifying the prior distribution as uniform with respect to the Hausdorff measure...
 \end{quote}

 But it is not immediately clear how one should approach the common scenario where the prior is defined \emph{a priori} using the embedding coordinates, i.e. those of the ambient Euclidean space. On the sphere, for example, such priors include the Von Mises-Fisher distribution. On the Stiefel manifold, such priors include the matrix Bingham-Von Mises-Fisher distribution \parencite{hoff2009simulation}.  Contrary to the above statement, one suspects that the log-Jacobian term should never be necessary, and this turns out to be the case. 

\section{Alternative derivation I}

Let $\pi(x)=\pi_\mathcal{H}(x)$ denote a target posterior density defined directly using embedding coordinates. For the unit sphere, this means that $x^Tx=1$; for the Stiefel manifold of $d\times s$ orthonormal matrices, this means that $x^Tx=I_s$, for $I_s$ the identity matrix of the given dimension $s$.   Let $\Pi_x$ be the the orthogonal projection onto the tangent space of the embedded manifold at point $x$. For example,  for the sphere, this projection is given by
\begin{align*}
    \Pi_x = I-xx^T \, ;
\end{align*}
for the Stiefel manifold, the matrix is (see Appendix \ref{note_appendix})
\begin{align*}
    \Pi_x = I_{ds} - \frac{1}{2} (I_{s^2} \otimes x) (P+I_{ds}) (I_{s^2} \otimes x^T)  \, ,
\end{align*}
for $\otimes$ the Kronecker product and $P$ the $ds\times ds$ permutation matrix for which $P \vecc(x)=\vecc(x^T)$ for any matrix $x$. For simplicity, we take the sphere as our prime example and leave the Stiefel manifold case for the appendix.

Let momentum $p$ follow a degenerate Gaussian distribution on the tangent space to the sphere at $x$, i.e. $p \sim N (0, \Pi_x M\,\Pi_x)$, where $M$ is some positive semi-definite matrix.  Then at any point $x$, the density of $p$ is proportional to
\begin{align*}
    \Det^{-1/2}(\Pi_xM\, \Pi_x) \exp\Big( -\frac{1}{2}p^T(\Pi_xM\, \Pi_x)^+p \Big)\, ,
\end{align*}
where $\Det(A)$ is the pseudo determinant and $A^+$ is the pseudo inverse of matrix $A$. Then the Hamiltonian is given by
\begin{align}\label{origHam}
    H(x,p) = - \log \pi(x) + \frac{1}{2}\log\, \Det(\Pi_xM\, \Pi_x)  + \frac{1}{2} p^T(\Pi_xM\, \Pi_x)^+p\, ,
\end{align}
for any pair $x$ and $p$. Similar to the original gMC algorithm, we split $H(x,p)$ into two Hamiltonians
\begin{align*}
    H^{[1]}(x,p) = - \log \pi(x) + \frac{1}{2}\log\, \Det(\Pi_xM\, \Pi_x) 
\end{align*}
and
\begin{align*}
    H^{[2]}(x,p) =  \frac{1}{2} p^T(\Pi_xM\, \Pi_x)^+p  \, .
\end{align*}
Using some matrix calculus (Appendix \ref{appC}) and the fact that $\nabla \Det(A)=\Det(A)\,A^+$  \parencite{holbrook2018differentiating}, the first system gives the equations
\begin{align}\label{first_system}
    \dot{x} &= 0 \\ \nonumber
    \dot{p} &= \nabla_x \log \pi(x) -  (\Pi_xM\, \Pi_x)^+\Pi_x M x \,.
\end{align}
Since the gradient $\nabla_x \log \pi(x)$ does not necessarily belong to the tangent space, we perform the change of variables $v= (\Pi_xM\, \Pi_x)^+p$. The  equations now read
\begin{align}\label{H1eqs}
    \dot{x} &= 0 \\ \nonumber
    \dot{v} &= (\Pi_xM\, \Pi_x)^+ \big( \nabla_x \log \pi(x) - (\Pi_xM\, \Pi_x)^+\Pi_x M x \big)\,.
\end{align}
Velocity $v$ stays on the tangent space at $x$ because $(\Pi_xM\, \Pi_x)^+=\Pi_x (\Pi_xM\, \Pi_x)^+\Pi_x$ in general.  The second system may also be rewritten:
\begin{align*}
    H^{[2]}(x,p) &= \frac{1}{2} p^T (\Pi_xM\, \Pi_x)^+ p \\ \nonumber
    &= \frac{1}{2} p^T(\Pi_xM\, \Pi_x)^+(\Pi_xM\, \Pi_x) (\Pi_xM\, \Pi_x)^+ p \\ \nonumber
    &= \frac{1}{2} v^T (\Pi_xM\, \Pi_x) v \\ \nonumber
    &= \frac{1}{2} \tilde{v}^T\tilde{v} := H^{[2]}(x,\tilde{v}) \, ,
\end{align*}
where $\tilde{v}=(\Pi_xM\Pi_x)^{1/2}v$. The system corresponding to $H^{[2]}$ is solved by the geodesic with initial conditions $(x,\tilde{v})$.  Thus the system corresponding to $H$ may be integrated by iteratively advancing according to \eqref{H1eqs} and spherical geodesics, alternating between $v$ and $\tilde{v}$ between steps. The general algorithm is given in Appendix A.1.

Accounting for the deterministic maps $v \mapsto \tilde{v}$ and $\tilde{v} \mapsto v$ within the accept/reject step yields a surprisingly simple acceptance probability. For the trajectory beginning at point $x_0$, $v$ is mapped to $\tilde{v}=(\Pi_{x_0}M\Pi_{x_0})^{1/2}v$ before the geodesic flow, but $\tilde{v}$ is mapped to $v=(\Pi_{x_1}M\Pi_{x_1})^{-1/2}\tilde{v}$ afterward, where we have used the shorthand $(\Pi_{x_1}M\Pi_{x_1})^{-1/2} = \left((\Pi_{x_1}M\Pi_{x_1})^+\right)^{1/2}$. But before the next geodesic flow, we apply the inverse map $v \mapsto \tilde{v}=(\Pi_{x_1}M\Pi_{x_1})^{1/2}v$. In this way, all internal deterministic maps cancel out, and one must only account for the first and last. Thus, for a trajectory consisting of $T$ steps, the Jacobian correction is
\begin{align*}
    \Det \left((\Pi_{x_0}M\, \Pi_{x_0})^{1/2}  (\Pi_{x_T}M\, \Pi_{x_T})^{-1/2}\right) \, ,
\end{align*}
and the resulting log acceptance probability is the minimum of 0 and
\begin{align}\label{alpha}
    \alpha &= - \log \pi (x_0) + \log \Det (\Pi_{x_0}M\, \Pi_{x_0}) +   \frac{1}{2}v_0^T(\Pi_{x_0}M\, \Pi_{x_0})v_0 + \\ \nonumber
    & \log \pi (x_T) - \log \Det (\Pi_{x_T}M\, \Pi_{x_T}) -  \frac{1}{2}v_T^T(\Pi_{x_T}M\, \Pi_{x_T})v_T \, .
\end{align}
See Appendix A.1 for details.

\section{Alternative derivation II}

But why begin with the momentum at all? By beginning with velocity, one may derive yet another class of algorithms that nonetheless reduces to the original geodesic Monte Carlo algorithm. The following approach is similar to that of \textcite{holbrook2017geodesic} and is related to the Lagrangian formulation found in \textcite{lan2015markov}.  We let the velocity have the same distribution as before, i.e. $v \sim N\big(0,(\Pi_xM\, \Pi_x)^+\big)$, and define the non-canonical (cf. \textcite{beskos2011hybrid}) Hamiltonian 
\begin{align*}
    H(x,v) = - \log \pi(x) - \frac{1}{2}\log\, \Det(\Pi_xM\, \Pi_x)  + \frac{1}{2} v^T(\Pi_xM\, \Pi_x)v\, .
\end{align*}
Note that the sign of the log pseudo determinant differs from that from Equation \eqref{origHam}, but the quadratic terms are equal. Again, split the Hamiltonian in two:
\begin{align*}
    H^{[1]}(x,v) = - \log \pi(x) - \frac{1}{2}\log\, \Det(\Pi_xM\, \Pi_x) \, , \quad \quad H^{[2]}(x,v) = \frac{1}{2} v^T(\Pi_xM\, \Pi_x)v \, .
\end{align*}
The first yields the equations
\begin{align*}
    \dot{x} &= 0 \\ \nonumber
    \dot{v} &= (\Pi_xM\, \Pi_x)^+ \big( \nabla_x \log \pi(x) + (\Pi_xM\, \Pi_x)^+\Pi_x M x \big)\,,
\end{align*}
where the only difference with Equation \eqref{H1eqs} is the sign of $(\Pi_xM\, \Pi_x)^+\Pi_x M x$.  The second Hamiltonian is handled in the exact same way as above.  Map $v \mapsto \tilde{v}=(\Pi_{x}M\Pi_{x})^{1/2}$, advance along the geodesics, and map back to $v=(\Pi_{x}M\Pi_{x})^{-1/2}\tilde{v}$.  As above, the same Jacobian correction appears in the accept/reject step, and this time the log acceptance probability simplifies even further (see Appendix A.2) to
\begin{align}\label{alpha2}
    \alpha &= - \log \pi (x_0) +   \frac{1}{2}v_0^T(\Pi_{x_0}M\, \Pi_{x_0})v_0 + \log \pi (x_T) -  \frac{1}{2}v_T^T(\Pi_{x_T}M\, \Pi_{x_T})v_T \, ,
\end{align}
i.e., the log pseudo determinants cancel. See Appendix A.2 for algorithmic details.

\section{Obtaining the original algorithm}
For both alternative derivations, the formulas greatly simplify when $M$ is the identity matrix, and the original geodesic Monte Carlo algorithm is obtained. Because the pseudo determinant of a projection matrix is unity, the Hamiltonians reduce to 
\begin{align*}
    H(x,v) = - \log \pi(x) \pm \frac{1}{2} \log \Det (\Pi_x) + \frac{1}{2} v^T\Pi_xv = - \log \pi(x) + \frac{1}{2} v^Tv\, .
\end{align*}
The simplified Hamiltonian is the same as Formula \eqref{gMCham}, but with $\pi(x)$ replacing $\pi_{\mathcal{H}}(x)$, the posterior with respect to the Hausdorff measure.  As established above, the two are equivalent, but by working completely with embedding coordinates, we are able to avoid any notion of intrinsic geometry whatsoever and thus require less mathematical machinery.

Similarly, there is no need for the Jacobian correction within the accept/reject step. Concretely, this is because $\tilde{v}=\Pi_x^{1/2}v=\Pi_xv=v$. Theoretically, this is because the geodesic Monte Carlo algorithm is not symplectic for general $M$ but is symplectic for $M$ the identity. Finally, the two derivations may be viewed as constructing random walks on the cotangent and tangent bundles, respectively.  The upshot is that the original geodesic Monte Carlo algorithm may be interpreted either way.

\section{Discussion}

We have proposed two alternative derivations of the geodesic Monte Carlo for embedded manifolds \parencite{byrne2013geodesic}.  These derivations are conceptually simpler, as they do not rely on a notion of intrinsic manifold geometry. They clarify the original algorithm by showing that  the  inclusion of the log-Jacobian of the embedding in the Hamiltonian is unnecessary in any case where the target distribution is defined using embedding coordinates. This claim goes beyond the statement of the original paper. 

Here, the original geodesic Monte Carlo algorithm was presented as a special case of two general classes of algorithms with non-trivial mass matrices. As a result, the new derivations emphasized the role played by the degenerate Gaussian distribution.  Finally, the exposition hinted how Metropolis adjustments may be incorporated into geometric Langevin algorithms such as \textcite{leimkuhler2016efficient}.

\printbibliography

\appendix

\section{Acceptance probabilities and generalized algorithms}\label{appA}
\subsection{First alternative derivation}

 Let $x_0$ be the trajectory's starting position and $x_T$ be its end point. Also let $h_0$ and $h_T$ denote the energies at the beginning and end of the trajectory, respectively.  Then the log acceptance probability is min$\left(0, \alpha \right)$, where
\begin{align*}
    \alpha &= h_0 - h_T + \frac{1}{2} \log \Det (\Pi_{x_0}M\, \Pi_{x_0}) - \frac{1}{2} \log \Det (\Pi_{x_T}M\, \Pi_{x_T}) \\
    &= - \log \pi (x_0) + \frac{1}{2} \log \Det (\Pi_{x_0}M\, \Pi_{x_0}) +  \frac{1}{2}v_0^T(\Pi_{x_0}M\, \Pi_{x_0})v_0 + \\
    & \log \pi (x_T) - \frac{1}{2} \log \Det (\Pi_{x_T}M\, \Pi_{x_T}) -  \frac{1}{2}v_T^T(\Pi_{x_T}M\, \Pi_{x_T})v_T + \\
    & \frac{1}{2} \log \Det (\Pi_{x_0}M\, \Pi_{x_0}) - \frac{1}{2} \log \Det (\Pi_{x_T}M\, \Pi_{x_T}) \\
    &= - \log \pi (x_0) + \log \Det (\Pi_{x_0}M\, \Pi_{x_0})+ \frac{1}{2}v_0^T(\Pi_{x_0}M\, \Pi_{x_0})v_0 + \\ 
    &\log \pi (x_T)   - \log \Det (\Pi_{x_T}M\, \Pi_{x_T})- \frac{1}{2}v_T^T(\Pi_{x_T}M\, \Pi_{x_T})v_T  \\
    &:= e_0 -e_T \, .
\end{align*}
In the final line, $e_0$ and $e_T$ denote the terms collected into those featuring the initial and final positions, respectively.

\begin{algorithm}[H]
  \caption{Embedding geodesic Monte Carlo with non-trivial mass matrix 1} \label{alg::gMC} 
  \raggedright{Let $x=x^{(k)}$ be the $k$th state of the Markov chain. The next sample is generated according to the following procedure.\\
  (a) Generate proposal state $x^*$:}
  \begin{algorithmic}[1]
   \State $v \sim N\big(0,(\Pi_xM\, \Pi_x)^+\big)$
   \State $e \gets - \log \pi (x)  +  \log \Det (\Pi_x M\, \Pi_x) +  \frac{1}{2}v^T(\Pi_xM\, \Pi_x) v$
   \State $x^* \gets x$
   \For{$\tau = 1,\dots, T$}
   	\State $v \gets v + \frac{\epsilon}{2} (\Pi_{x^*}M\, \Pi_{x^*})^+ \big( \nabla_{x^*} \log \pi(x^*) - (\Pi_{x^*}M\, \Pi_{x^*})^+\Pi_{x^*} M x^* \big)$
   	\State $\tilde{v} \gets (\Pi_{x^*}M\Pi_{x^*})^{1/2}v$
	\State \parbox[t]{\dimexpr\linewidth-3em}{Progress $(x^*,\tilde{v})$ along the geodesic flow for time $\epsilon$.\strut}
	\State $v \gets (\Pi_{x^*}M\Pi_{x^*})^{-1/2}\tilde{v}$
   	\State $v \gets v + \frac{\epsilon}{2} (\Pi_{x^*}M\, \Pi_{x^*})^+ \big( \nabla_{x^*} \log \pi(x^*) - (\Pi_{x^*}M\, \Pi_{x^*})^+\Pi_{x^*} M x^* \big)$
   \EndFor
   \State $e^* \gets - \log \pi (x^*) +   \log \Det (\Pi_{x^*} M\, \Pi_{x^*})  +  \frac{1}{2}v^T(\Pi_{x^*}M\, \Pi_{x^*}) v$
   \end{algorithmic}
  \raggedright{(b) Accept proposal with log probability $\min\{1, \exp(e)/\exp(e^*) \} $:}
    \begin{algorithmic}[1]
    \State $u \sim U(0, 1)$
    \If{$u < \exp (e-e^*)$ }
    	\State $x \gets x^*$
    \EndIf
  \end{algorithmic}
   \raggedright{(c) Assign value $x$ to $x^{(k+1)}$, the $(k+1)$th state of the Markov chain.}
\end{algorithm}

\subsection{Second alternative derivation}

Again let $x_0$ be the trajectory's starting position and $x_T$ be its end point. Let $e_0$ and $e_T$ denote the energies at the beginning and end of the trajectory, respectively.  Then the log acceptance probability is min$\left(0, \alpha \right)$, where
\begin{align*}
    \alpha &= h_0 - h_T + \frac{1}{2} \log \Det (\Pi_{x_0}M\, \Pi_{x_0}) - \frac{1}{2} \log \Det (\Pi_{x_T}M\, \Pi_{x_T}) \\
    &= - \log \pi (x_0) - \frac{1}{2} \log \Det (\Pi_{x_0}M\, \Pi_{x_0}) +  \frac{1}{2}v_0^T(\Pi_{x_0}M\, \Pi_{x_0})v_0 + \\
    & \log \pi (x_T) + \frac{1}{2} \log \Det (\Pi_{x_T}M\, \Pi_{x_T}) -  \frac{1}{2}v_T^T(\Pi_{x_T}M\, \Pi_{x_T})v_T + \\
    & \frac{1}{2} \log \Det (\Pi_{x_0}M\, \Pi_{x_0}) - \frac{1}{2} \log \Det (\Pi_{x_T}M\, \Pi_{x_T}) \\
    &= - \log \pi (x_0) + \frac{1}{2}v_0^T(\Pi_{x_0}M\, \Pi_{x_0})v_0 +\log \pi (x_T)  - \frac{1}{2}v_T^T(\Pi_{x_T}M\, \Pi_{x_T})v_T  \\
    &:= e_0 -e_T \, .
\end{align*}
In the final line, $e_0$ and $e_T$ denote the terms collected into those featuring the initial and final positions, respectively.

\begin{algorithm}[H]
  \caption{Embedding geodesic Monte Carlo with non-trivial mass matrix 2} \label{alg::gMC2} 
  \raggedright{Let $x=x^{(k)}$ be the $k$th state of the Markov chain. The next sample is generated according to the following procedure.\\
  (a) Generate proposal state $x^*$:}
  \begin{algorithmic}[1]
   \State $v \sim N\big(0,(\Pi_xM\, \Pi_x)^+\big)$
   \State $e \gets - \log \pi (x) +  \frac{1}{2}v^T(\Pi_xM\, \Pi_x) v$
   \State $x^* \gets x$
   \For{$\tau = 1,\dots, T$}
   	\State $v \gets v + \frac{\epsilon}{2} (\Pi_{x^*}M\, \Pi_{x^*})^+ \big( \nabla_{x^*} \log \pi(x^*) + (\Pi_{x^*}M\, \Pi_{x^*})^+\Pi_{x^*} M x^* \big)$
   	\State $\tilde{v} \gets (\Pi_{x^*}M\Pi_{x^*})^{1/2}v$
	\State \parbox[t]{\dimexpr\linewidth-3em}{Progress $(x^*,\tilde{v})$ along the geodesic flow for time $\epsilon$.\strut}
	\State $v \gets (\Pi_{x^*}M\Pi_{x^*})^{-1/2}\tilde{v}$
   	\State $v \gets v + \frac{\epsilon}{2} (\Pi_{x^*}M\, \Pi_{x^*})^+ \big( \nabla_{x^*} \log \pi(x^*) + (\Pi_{x^*}M\, \Pi_{x^*})^+\Pi_{x^*} M x^* \big)$
   \EndFor
   \State $e^* \gets - \log \pi (x^*)   +  \frac{1}{2}v^T(\Pi_{x^*}M\, \Pi_{x^*}) v$
   \end{algorithmic}
  \raggedright{(b) Accept proposal with log probability $\min\{1, \exp(e)/\exp(e^*) \} $:}
    \begin{algorithmic}[1]
    \State $u \sim U(0, 1)$
    \If{$u < \exp (e-e^*)$ }
    	\State $x \gets x^*$
    \EndIf
  \end{algorithmic}
   \raggedright{(c) Assign value $x$ to $x^{(k+1)}$, the $(k+1)$th state of the Markov chain.}
\end{algorithm}

\section{Projection matrix for the Stiefel manifold}\label{note_appendix}

When modeling an element $x \in \mathcal{S}(d,s)$ of the Stiefel manifold, for $d\times s$ momentum matrix we write the degenerate Gaussian distribution
\begin{align*}
    \Det^{-1/2}(\Pi_xM\, \Pi_x) \exp\Big( -\frac{1}{2}\vecc(p)^T(\Pi_xM\, \Pi_x)^+\vecc(p) \Big)\, ,
\end{align*}
$\Pi_x$ and $M$ are $ds\times ds$ matrices.  To get the form for $\Pi_x$, we note that the orthogonal projection of a matrix $v$ onto the tangent space at $x$ is
\begin{align*}
  \Pi_x(v) =  v - \frac{1}{2} x (v^Tx + x^Tv) \, .
\end{align*}
Applying the vec operator gives
\begin{align*}
  \vecc( \Pi_x(v)) &=  \vecc(v) - \frac{1}{2} \vecc \big( x (v^Tx + x^Tv) \big)  \\ 
                 &=  \vecc(v) - \frac{1}{2} (I_{s^2} \otimes x) \vecc(v^Tx + x^Tv) \\
                 &=  \vecc(v) - \frac{1}{2} (I_{s^2} \otimes x) \vecc(v^Tx) + \vecc(x^Tv) \\
                 &=  \vecc(v) - \frac{1}{2} (I_{s^2} \otimes x) P\, \vecc(x^T) + x^Tv) \\ 
                 &=  \vecc(v) - \frac{1}{2} (I_{s^2} \otimes x) (P+I_{ds}) \vecc(x^Tv) \\
                 &=  \vecc(v) - \frac{1}{2} (I_{s^2} \otimes x) (P+I_{ds}) (I_{s^2} \otimes x^T) \vecc(v) \\ 
                 &=  \big(I_{ds} - \frac{1}{2} (I_{s^2} \otimes x) (P+I_{ds}) (I_{s^2} \otimes x^T)\big) \vecc(v) \\ 
                 &= \Pi_x \, v
\end{align*}
Hence 
\begin{align*}
    \Pi_x = I_{ds} - \frac{1}{2} (I_{s^2} \otimes x) (P+I_{ds}) (I_{s^2} \otimes x^T) \, .
\end{align*}

\section{Deriving the first system of equations}\label{appC}

To obtain Equation \eqref{first_system}, we need to calculate
\begin{align*}
    \nabla_x \log \Det (\Pi_x M \Pi_x)\, .
\end{align*}
This may be done using the differential and Theorem 2.20 from \textcite{holbrook2018differentiating}, namely
\begin{align*}
    d \Det(A) = \Det(A) \mbox{tr} \left(A^+ (dA) \right) \, .
\end{align*}
Thus
\begin{align*}
    d \log \Det (\Pi_x M \Pi_x) &=  \mbox{tr} \left((\Pi_x M \Pi_x)^+ \left(d(\Pi_x M \Pi_x)\right) \right) \\
    &= \mbox{tr} \left((\Pi_x M \Pi_x)^+ \left((d\Pi_x) M \Pi_x + \Pi_x M (d\Pi_x))\right) \right) \, .
\end{align*}
$d\Pi_x$ is given by
\begin{align*}
    d\Pi_x &= d(I-xx^T) =- d(xx^T) \\
    &= - (dx)x^T - x(dx)^T \, ,
\end{align*}
so we have
\begin{align*}
    d \log \Det (\Pi_x M \Pi_x) &= - \mbox{tr} \left((\Pi_x M \Pi_x)^+ \left(\left((dx)x^T + x(dx)^T\right) M \Pi_x  + \Pi_x M \left((dx)x^T + x(dx)^T \right)\right) \right) \, .
\end{align*}
Distributing the leading $(\Pi_x M \Pi_x)^+$ and rearranging terms gives
\begin{align*}
    d \log \Det (\Pi_x M \Pi_x) &= - 2\, \left( (dx)^T \left(M\Pi_x(\Pi_x M \Pi_x)^+x + (\Pi_x M \Pi_x)^+\Pi_xMx  \right) \right) \, ,
\end{align*}
but the first term of the inner parenthesis is equal to zero because $(\Pi_x M \Pi_x)^+ = \Pi_x(\Pi_x M \Pi_x)^+ \Pi_x$ and so
\begin{align*}
    M\Pi_x(\Pi_x M \Pi_x)^+x &= M\Pi_x(\Pi_x M \Pi_x)^+ \Pi_xx \\
    &= M\Pi_x(\Pi_x M \Pi_x)^+ 0 = 0 \, .
\end{align*}
Hence,
\begin{align*}
    d \log \Det (\Pi_x M \Pi_x) &= - 2\,  \left( (dx)^T (\Pi_x M \Pi_x)^+\Pi_xMx \right) \, ,
\end{align*}
and it follows immediately that
\begin{align*}
    \nabla_x \log \Det (\Pi_x M \Pi_x) =-2\, (\Pi_x M \Pi_x)^+\Pi_xMx \, .
\end{align*}
\end{document}